\documentclass{amsart}
\usepackage{ae,aecompl}\usepackage{ifthen}
\usepackage[a4paper,twoside]{geometry}
\usepackage{amsmath, amsthm}
\usepackage{graphicx}
\usepackage{amssymb}
\usepackage{url}
\newboolean{dum}
\setboolean{dum}{false}
\newcommand{\comment}[1]{\ifthenelse{\boolean{dum}}{
{\par\noindent\Huge\ding{46}} \fbox{\parbox{10cm}{#1}}\par}{}}

\newtheorem{lemma}[subsection]{Lemma}

\newtheorem{prop}{Proposition}
\newtheorem{cor}[subsection]{Corollary}

\numberwithin{equation}{section}

\begin{document}
\title{Edge effects in some perturbations of the GUE}
\author{K.E.~Bassler}
\address{
Department of Physics,
University of Houston,
Houston, USA}
\author{P.J. Forrester}
\address{Department of Mathematics and Statistics,
The University of Melbourne,
Victoria 3010, Australia}
\author{N.E.~Frankel}
\address{School of Physics,
The University of Melbourne,
Victoria 3010, Australia}
\begin{abstract}{A bordering of GUE matrices is considered, in which the bordered row consists of
zero mean complex Gaussians N$[0,\sigma/2] + i {\rm N}[0,\sigma/2]$ off the diagonal, and the real Gaussian
N$[\mu,\sigma/\sqrt{2}]$ on the diagonal. We compute the explicit form of the eigenvalue probability function
for such matrices, as well as that for matrices obtained by repeating the bordering. The correlations are
in general determinantal, and in the single bordering case the explicit form of the correlation kernel is
computed. In the large $N$ limit it is shown that $\mu$ and/or $\sigma$ can be tuned to induce a separation of the
largest eigenvalue. This effect is shown to be controlled by a single parameter, universal correlation kernel.
}
\end{abstract}
\maketitle

\section{Introduction}
As is well known, in what was the first application of random matrix theory to physics,
Wigner introduced a particular ensemble of real symmetric matrices as a model of the
highly excited states of complex nuclei (see e.g.~\cite{Po65}). Thus the features of the
Hamiltonian hypothesized to be responsible for the distribution of these states---in a 
minimalist reduction chosen to be simply time reversal symmetry and the absence of a
preferential basis---were observed to be features shared by the ensemble of  real symmetric
matrices with
probability density function
(PDF) proportional to the Gaussian $\exp(-{\rm Tr} \, X^2/2)$. Hermitian matrices of size $N \times N$,
with $N$ large (formally $N \to \infty$) are chosen because the discrete portion of the spectrum
consisting of an infinite number of levels are being modelled. The presence of a time reversal
symmetry means the basis elements of the Hamiltonian and thus the Hermitian matrix can be chosen to
be real, thus explaining the use of real symmetric matrices, and the PDF is the simplest that is
invariant under mappings $X \mapsto R X R^T$ where $R$ is a real orthogonal matrix, this having the
physical interpretation of there being no preferential basis. This invariance lends the name
Gaussian orthogonal ensemble (GOE) to this class of random real symmetric matrices.

Of direct relevance to the concerns of the present paper is a variant of the GOE specified by the
PDF  proportional to $\exp(-{\rm Tr} \, (X - X_0)^2/2)$. Here $X_0$ is a fixed real symmetric matrix
specifying the mean of the Gaussian distribution. Scaling this PDF by a parameter $t$ so that it reads
$\exp(-{\rm Tr} \, (X - X_0)^2/2t)$, one obtains a well known (see e.g.~\cite{Ha90,Fo10})
realization of the Dyson Brownian motion model \cite{Dy62b}. In this the corresponding eigenvalue PDF
evolves in $t$ according to a certain Fokker-Planck equation, with the initial condition that the
eigenvalue PDF is a product of delta functions at the eigenvalues of $X_0$.

Shifting the mean in the PDF for the GOE is equivalent to forming an ensemble of matrices
of the form $X + X_0$ where $X$ is a member of the GOE and $X_0$ is as above. Special to this
class of matrices, and a primary concern of our work herein, is the case that $X_0$ is of low rank.
The simplest of these is when $X_0 = \varepsilon \vec{e}_1 \vec{e}_1^T$, where $\vec{e}_1$ is the elementary
column vector with 1 in the first entry. By orthogonal invariance of the distribution of $X$ the
corresponding ensemble of matrices has identical spectral properties to the shifted mean GOE ensemble
 in which $X_0$ has all entries equal to $\varepsilon$. This ensemble appeared in an analysis of a
spherical spin glass due to Kosterlitz, Thouless and Jones \cite{KTJ78} (see also 
\cite{AR10}), and it has also been used as a model Hamiltonian in the study of mesoscopic
quantum structures \cite{SS03}.

Tuning the parameter $\varepsilon$ by writing $\varepsilon = c / \sqrt{2N}$, $(c > 0)$, it was shown in
\cite{KTJ78} that in in the large $N$ limit the largest eigenvalue undergoes a phase
transition as a function of $c$. Thus for $0 < c < 1$ the leading order location of the
largest eigenvalue is at $\sqrt{2N}$ (which is the edge of the support of the spectrum for
$\varepsilon = 0$), while for $c > 1$ it is at 
\begin{equation}\label{c}
(N/2)^{1/2}(c + 1/c) 
\end{equation}
and thus separates from the leading support of the spectrum. The same effect occurs for the
ensemble of complex Hermitian Gaussian matrices with PDF proportional to
$\exp(-{\rm Tr} \, X^2)$. This specifies the Gaussian unitary ensemble (GUE). Moreover, is this
case the analytic form of the scaled correlation functions can be computed exactly
\cite{Pe06,DF06}. 

In this work we will analyze the scaling regime of this eigenvalue separation effect
for a generalization of the additive perturbation $X + X_0$, $X$ a member of the
GUE and $X_0$ of low rank. The main generalization to be considered is the class of bordered
matrices
\begin{equation}\label{1}
\begin{bmatrix} c_N & \vec{v}_N^\dagger \\ \vec{v}_N & G_N \end{bmatrix}
\end{equation}
where $G_N$ a member of the GUE, $c_N \mathop{=}\limits^{\rm d}  {\rm N}[\mu,{\sigma}/\sqrt{2}]$
and $v_N$ a $N \times 1$ vector of complex
Gaussians ${\rm N}[0,\sigma/2] + i {\rm N}[0,\sigma/2]$. 
In the case that $\sigma = 1$ this corresponds to perturbed GUE matrices
$X + {\rm diag} \, [\mu,0^r]$ with $X$ a member of the $(N+1) \times (N+1)$ GUE and the notation
$0^n$ denoting the eigenvalue 0 repeated $n$ times. Up to a similarity transformation, such matrices
are equivalent to $(N+1) \times (N+1)$ shifted mean GUE matrices $X + \varepsilon e_1 e_1^T$,
$\varepsilon = \mu/(N+1)$.
We will furthermore consider interations
of (\ref{1}) obtained by bordering it with a new first row and column with elements distributed
as in the previous first row and column. Thus we are generalizing the shifted mean GUE ensemble
by also allowing for a different variance along the first row and first column, or more
generally the first $r$ rows and columns.

In a previous paper \cite{BFF09} we gave an extended discussion of the relevance of matrix
ensembles with varying mean and variance to contemporary studies in mathematical statistics and
applied mathematics. Most notably these include the study of spiked models in the analysis of
multivariate data \cite{BBP05}, and stability questions relating to ecological webs
\cite{Ma72}. Some universal behaviour has been found. Thus it has been shown \cite{Pe06,DF06}
that the critical region of the $r=1$ complex spiked model (complex Wishart matrices with a
covariance matrix having a single eigenvalue different from 1) is the same as that for the
rank 1 shifted mean GUE. We will exhibit the same critical correlations  for the
model (\ref{1}) as a function of $\sigma$ in the case $\mu = 0$, or more generally as a function
of both $\mu$ and $\sigma$, when tuned about the eigenvalue separation point.

Our first task, addressed in Section 2, is to give a determinantal formula for the
eigenvalue PDF of matrices (\ref{1}) and their extensions defined by bordering further rows
and columns. The determinantal structure is used in Section 3 to compute the correlation
functions for fixed values of the parameters. With the parameters tuned in the neighbourhood
of the eigenvalue separation point, the $N \to \infty$ scaling limit is studied in Section 4.

\section{Joint eigenvalue PDF}
We begin by observing that the eigenvalue PDF for matrices of the form (\ref{1}) is unchanged
by the replacement of $G_N$ by diag$\, G_N$.

\begin{lemma}\label{lem1}
Consider matrices of the form (\ref{1}), generalized so that $G_N$ is any Hermitian matrix independent
of $c_N$ and $v_N$. Then (\ref{1}) and the random matrices
\begin{equation}\label{2}
\begin{bmatrix} c_N & \vec{v}_N^\dagger \\ \vec{v}_N & {\rm diag} \,{G}_N \end{bmatrix}
\end{equation}
have the same eigenvalue PDF.
\end{lemma}

\noindent
{\it Proof}. Since $G_N$ is assumed Hermitian we can write $U^\dagger G_N U = {\rm diag} \, G_N$ for
some $N \times N$ unitary matrix $U$. Noting too that a complex Gaussian is unchanged in distribution
when acted on by a unitary matrix we see that (\ref{2}) has the same distribution as 
$$
\begin{bmatrix} 1 & \vec{0}_N^\dagger \\ \vec{0}_N & U^\dagger \end{bmatrix}
\begin{bmatrix} c_N & \vec{v}_N^\dagger \\ \vec{v}_N & G_N \end{bmatrix}
\begin{bmatrix} 1 & \vec{0}_N^\dagger \\ \vec{0}_N & U \end{bmatrix}.
$$
This implies the result. \hfill $\square$

\medskip

Suppose
\begin{equation}\label{3a}
{\rm diag} \, G_N = {\rm diag} \, [a_1,a_2,\dots,a_N]
\end{equation}
and let us denote by $\chi(\lambda \succ a)$ the indicator function for the interlaced region
\begin{equation}\label{4}
\lambda_1 > a_1 > \lambda_2 > a_2 > \cdots a_N > \lambda_{N+1}.
\end{equation}
We can use theory from \cite{FR02b} to give the explicit form of the eigenvalue PDF
of (\ref{2}) for given $\{a_i\}$.

\begin{prop}\label{p2}
Consider random matrices (\ref{2}), with diag$\, G_N$ given by (\ref{3a}) and $c_N$,
$\vec{v}_N$ having Gaussian distribution specified below (\ref{1}). With $\{a_i\}$ given, the
eigenvalue PDF is
\begin{eqnarray}\label{3}
&&\Big ( {\sqrt{2} \over \sigma} \Big )^{2N+2} \sqrt{1 \over 2 \pi }
e^{- \sigma^{-2}
 \sum_{j=1}^{N+1} \lambda_j^2 - (1 - \sigma^{-2}) \sum_{j=1}^N a_j^2 }
e^{-(\mu/\sigma)^2} \nonumber \\
&& \qquad \times e^{2 \mu \sigma^{-2} ( \sum_{j=1}^{N+1} \lambda_j -
\sum_{j=1}^N a_j)}
{\prod_{1 \le j < k \le N + 1} (\lambda_j - \lambda_k) \over
\prod_{1 \le j < k \le N} (a_j - a_k) } \chi(\lambda \succ a).
\end{eqnarray}
\end{prop}

\noindent
{\it Proof.} \quad
Elementary manipulation of the corresponding
characteristic polynomial corresponding to (\ref{2})
shows that the condition for $\lambda$ to be 
an eigenvalue is 
\begin{equation}\label{3.1}
0 = \lambda - c_N - \sum_{j=1}^N {|v_N^{(j)}|^2 \over \lambda - a_j},
\end{equation}
where $v_N^{(j)}$ denotes the $j$th component of $\vec{v}_N$.
 By the
specification of the latter given below (\ref{1}) we see that
$$
|v_N^{(j)}|^2  \mathop{=}\limits^{\rm d} \Gamma[1,\sigma^2/2]
 \mathop{=}\limits^{\rm d} (\sigma^2/2)  \Gamma[1,1],
$$
where $\Gamma[k,c]$ refers to the gamma distribution, specified by the
PDF proportional to $(s/c)^{k-1} e^{-s/c}$. Writing
$$
{\lambda} = \Big ( {\sigma \over \sqrt{2}} \Big ) \tilde{\lambda}, \quad
c_N =  \Big ( {\sigma \over \sqrt{2}} \Big ) \tilde{c}, \quad
a_j =  \Big ( {\sigma \over \sqrt{2}} \Big ) \tilde{a}_j
$$
then allows (\ref{3.1}) to be written
\begin{equation}\label{3.3}
0 = \tilde{\lambda} - \tilde{c} - \sum_{j=1}^N {w_j \over \tilde{\lambda} -
\tilde{a}_j},
\end{equation}
where $w_j \mathop{=}\limits^{\rm d} \Gamma[1,1]$.

The distribution of the zeros of the zeros of the
random rational function (\ref{3.3}), and thus the
eigenvalues $\{\lambda_j\}_{j=1,\dots,N}$ in the case of $c_N$ fixed, is given by
\cite{FR02b} (Prop.~8 with $n=N$, $s_1 = \cdots = s_N = 1$,
$\lambda \mapsto \tilde{\lambda} - \tilde{c}$, $a_j \mapsto \tilde{a}_j - \tilde{c}$).
Explicitly, in terms of
the original variables $\{a_j\}$, $\{\lambda_j\}$ we have that the PDF
for the zeros is supported on the region $\chi(\lambda \succ a)$ and is
subject to the further constraint
\begin{equation}\label{3.5}
\sum_{l=1}^{N+1} \lambda_l = \sum_{l=1}^N a_l + c_N.
\end{equation}
In this region the PDF has the explicit functional form
\begin{equation}\label{3.6}
\Big ( {\sqrt{2} \over \sigma } \Big )^{2N+1}
{\prod_{1 \le j < k \le N + 1} (\lambda_j - \lambda_k) \over
\prod_{1 \le j < k \le N} (a_j - a_k) }
e^{- \sigma^{-2} ( \sum_{j=1}^{N+1} \lambda_j^2 - \sum_{j=1}^N a_j^2 )} e^{\sigma^{-2} c_N^2}.
\end{equation}
In the case that $c  \mathop{=}\limits^{\rm d}  {\rm N}[\mu,{\sigma}/\sqrt{2}]$, the reasoning
of \cite[Corollary 4]{FR02b} tells us that the constraint (\ref{3.5}) should be removed,
and in (\ref{3.6}) the term $\exp(-\sigma^{-2} c_N^2)$ should be replaced by
\begin{equation}\label{r4}
\sqrt{1 \over \pi \tilde{\sigma}^2}
e^{2 \mu \tilde{\sigma}^{-2} ( \sum_{l=1}^{N+1} \lambda_l - \sum_{l=1}^N a_l)}
e^{- \tilde{\sigma}^{-2} \mu^2}.
\end{equation}
This gives (\ref{3}). \hfill $\square$

\medskip
Consider now a random Hermitian matrix $G_{N+r}$ specified so that the first $r$ diagonal entries have
distribution N$[\mu,\sigma/\sqrt{2}]$, the entries to the right of these rows have
distribution N$[0,\sigma/2] + i {\rm N}[0,\sigma/2]$, while the bottom $N \times N$ sub-block is a
GUE matrix. According to Lemma \ref{lem1}, we have the inductive construction
\begin{equation}\label{GG}
G_{N+s} =
\begin{bmatrix} c_{N+s-1} & \vec{v}_{N+s-1}^\dagger \\
\vec{v}_{N+s-1} & G_{N+s-1} \end{bmatrix} \qquad (s=1,\dots,r).
\end{equation}
We can use Proposition \ref{p2} to give the joint PDF of the eigenvalues
$\{ \lambda^{(n+2)} \}_{s=0,1,\dots,r}$, $\lambda^{(j)} := (\lambda_1,\dots,\lambda_N)$.

\begin{prop}
Let $G_N$ be a member of the GUE, and let $G_{N+s}$ $(s=1,\dots,r)$ be specified by (\ref{GG}).
The joint eigenvalue PDF of these matrices is given by
\begin{eqnarray}\label{GH}
&&
{1 \over C_N} {(2 \sigma^{-2})^{rN + r^2/2} \over (2 \pi)^{r/2} }
e^{- \sigma^{-2} \sum_{j=1}^{N+r} (\lambda_j^{(N+r)})^2 + 
2 \mu \sigma^{-2} \sum_{j=1}^{N+r} \lambda_j^{(N+r)} } \nonumber \\
&& \quad \times \prod_{1 \le j < k \le N + r} (\lambda_j^{(N+r)} - \lambda_k^{(N+r)} )
\prod_{s=1}^r \chi(\lambda^{(N+s)} \succ \lambda^{(N+s-1)}) \nonumber \\
&& \quad \times \prod_{1 \le j < k \le N} (x_j^{(N)} - x_k^{(N)})
e^{( -1 + \sigma^{-2} ) \sum_{j=1}^{N} (\lambda_j^{(N)})^2 
-2 \mu \sigma^{-2} \sum_{j=1}^{N} \lambda_j^{(N)} }
\end{eqnarray}
where 
\begin{equation}\label{CN}
C_N = \pi^{N/2} 2^{-N(N-1)/2} \prod_{j=0}^N j!.
\end{equation}
With $\{q_j(x) \}_{j=0,1,\dots}$, $\{p_j(x)\}_{j=0,1,\dots}$ two sets of arbitrary
monic polynomials labelled by their degree, this can be rewritten to read
\begin{eqnarray}\label{GH1}
&&
{1 \over \tilde{C}_N} {(2 \sigma^{-2})^{rN + r^2/2} \over (2 \pi)^{r/2} }
e^{- \sigma^{-2} \sum_{j=1}^{N+r} (\lambda_j^{(N+r)})^2 + 
2 \mu \sigma^{-2} \sum_{j=1}^{N+r} \lambda_j^{(N+r)} } \nonumber \\
&& \quad \times \det [q_{j-1}(\lambda_k^{(N+r)} ]_{j,k=1,\dots,N+r}
\prod_{s=1}^r \det [ \chi_{\lambda_j^{(N+s)} > \lambda_k^{(N+s-1)}} ]_{j,k=1,\dots,N+s} \nonumber \\
&& \quad \times \det [ p_{j-1}(\lambda_k^{(N)} ]_{j,k=1,\dots,N}
e^{( -1 + \sigma^{-2} ) \sum_{j=1}^{N} (\lambda_j^{(N)})^2     
-2 \mu \sigma^{-2} \sum_{j=1}^{N} \lambda_j^{(N)} }
\end{eqnarray}
where $\tilde{C}_N = (-1)^{N(N-1)/2 + (N+r)(N+r-1)/2} C_N$.
\end{prop}

\noindent
{\it Proof.} \quad The conditional eigenvalue PDF of $G_{N+s}$, given the eigenvalues of
$G_{N+s-1}$ is equal to (\ref{3}) with $\lambda = \lambda^{(N+s)}$, $a = \lambda^{(N+s-1)}$.
Forming the product over $s$ $(s=r,r-1,\dots,1)$ gives the joint eigenvalue PDF of
$\{G_{N+s}\}_{s=1,\dots,r}$ given the eigenvalues of $G_N$. But the PDF for the latter is equal to
\cite[Prop.~1.3.4 multiplied by $N!$ to account for the ordering of $\lambda^{(N)}$]{Fo10}
\begin{equation}\label{B}
{1 \over C_N} \prod_{j=1}^N e^{- (\lambda_j^{(N)})^2} 
\prod_{1 \le j < k \le N}(\lambda_k^{(N)} - \lambda_j^{(N)})^2,
\end{equation}
which we multiply this conditional probability by to get (\ref{GH}).
The form (\ref{GH1}) follows from (\ref{GH}) by noting that
\begin{equation}\label{G4}
\chi(\lambda \succ a) = \det [ \chi_{\lambda_j > a_k} ]_{j,k=1,\dots,N+1}
\end{equation}
valid for $a_1 > a_2 > \cdots > a_N > a_{N+1} := - \infty$ (see e.g.~\cite[Prop.~5.9.1]{Fo10}),
and that,
with $\{r_j(\lambda)\}_{j=0,\dots,N}$ a set  
of monic polynomials,
\begin{equation}\label{G5}
\prod_{1 \le j < k \le N} (a_j - a_k) =
(-1)^{N(N-1)/2} \det [ r_{j-1}(a_k) ]_{j,k=1,\dots,N}.
\end{equation}
which is a consequence of the Vandermonde
identity. \hfill $\square$

\medskip
Our present interest is in the distribution of $\lambda^{(N+r)}$ only. Thus we must integrate
(\ref{GH1}) over $\lambda^{(N+s)}$, $s=0,1,\dots,r-1$.

\begin{prop}\label{p3}
Let $G_{N+r}$ be defined by the recursive construction (\ref{GG}), and let 
$\lambda^{(N+r)}$ denote the $N + r$ ordered eigenvalues. The PDF of
$\lambda^{(N+r)}$ is equal to 
\begin{eqnarray}\label{GH2}
&&
{1 \over \tilde{C}_N \prod_{s=1}^{r-1} s!}
 {(2 \sigma^{-2})^{rN + r^2/2} \over (2 \pi)^{r/2} }
e^{- \sigma^{-2} \sum_{j=1}^{N+r} (\lambda_j^{(N+r)})^2 }
e^{2 \mu \sigma^{-2} \sum_{j=1}^{N+r} \lambda_j^{(N+r)} } \nonumber \\
&& \quad \times \det [q_{j-1}(\lambda_k^{(N+r)} ]_{j,k=1,\dots,N+r}
\det \Big [ [h_{k-1,r-1}(\lambda_j^{(N+r)})]_{j=1,\dots,N+r \atop k=1,\dots,N} \quad
[(\lambda_j^{(N+r)})^{r-s} ]_{j=1,\dots,N+r \atop s=1,\dots,r} \Big ]
\end{eqnarray}
where
\begin{equation}\label{h}
h_{k,r}(x) = {1 \over r!}
\int_{-\infty}^x (x - u)^r e^{(-1+\sigma^{-2}) u^2 - 2 \mu \sigma^{-2} u}
p_k(u) \, du.
\end{equation}
\end{prop}

\noindent
{\it Proof.} \quad We integrate over $\lambda^{(N)}, \lambda^{(N+1)},\dots,\lambda^{(N+r-1)}$
in order. First note that the integrand is symmetric in $\lambda^{(N)}$ so the 
corresponding integration can be taken over all $\mathbb R^N$ provided we divide by $N!$. It is also
true that both determinants involving $\lambda^{(N)}$ are individually anti-symmetric in
$\lambda^{(N)}$. One of these is $\det [p_{k-1}(\lambda_j^{(N)}]_{j,k=1,\dots,N}$. The latter property means
we can replace this by its diagonal term $N! \prod_{k=1}^N p_{k-1}(\lambda_j^{(N)})$.
The $\lambda^{(N)}$ dependent terms in the integrand are thus a product over factors involving
$\lambda_k^{(N)}$, times the other determinant. The former can therefore be multiplied into column
$k$ $(k=1,\dots,N)$ of the latter and the integration over each $\lambda_k^{(N)}$ can be done column-by-column
to give
\begin{equation}\label{H}
\det \Big [ [h_{k,0}(\lambda_j^{(N+1)}) ]_{j=1,\dots,N+1 \atop k=1,\dots,N} \quad
[1]_{j=1,\dots,N+1} \Big ].
\end{equation}
Here we have used the fact that $\chi_{\lambda_j^{(N+1)} > \lambda_k^{(N)}} = 1$ for $k=N+1$ since
$\lambda_{N+1}^{(N)} = - \infty$.

The quantity (\ref{H}) must be multiplied by
$$
\det \Big [ \chi_{\lambda_j^{(N+2)} > \lambda_k^{(N+1)}} \Big ]_{j,k=1,\dots,N + 2}
$$
and the variables $\lambda^{(N+1)}$ integrated over. We make use of an analogous strategy to that just used.
Thus we note that the integration over $\lambda^{(N+1)}$ can be taken over all $\mathbb R^{N+1}$
provided we divide by $(N+1)!$ and furthermore (\ref{H}) can be replaced by
$$
(N+1)! \prod_{k=1}^N h_{k,0}(\lambda_k^{(N+1)}).
$$
Multiplying this into the columns of the remaining $x^{(N+1)}$-dependent determinant, and
integrating over $\lambda^{(N+1)}$ column-by-column gives 
\begin{equation}\label{H1}
\det \bigg [ \Big [ \int_{-\infty}^{\lambda_j^{(N+2)}} h_{k,0}(v) \, dv \Big ]_{j=1,\dots,N+2 \atop
k=1,\dots,N} \: \: [x_j^{(N+2)}]_{j=1,\dots,N+2} \: \:
[1]_{j=1,\dots,N+2} \bigg ].
\end{equation}
Here, to obtain the second last column we have used $\int^x du = x + c$ and have
subtracted $c$ by using the fact that the final column is all $1$'s. Furthermore, the integral
in (\ref{H1}) can be simplified
$$
\int_{-\infty}^x h_{k,0}(v) \, dv = \int_{-\infty}^x {d \over d v} (v - x) h_{k,0}(v) \, dv =
h_{k,1}(x),
$$
where the second equality follows by integration by parts.

Continuing this procedure until we have integrated over all $\lambda^{(N)}, \lambda^{(N+1)}, \dots,
\lambda^{(N+r-1)}$ gives (\ref{GH2}) for the marginal distribution of $\lambda^{(N+r)}$.
\hfill $\square$

\medskip
Some checks on (\ref{GH2}) are possible. Thus the case $N=0$ corresponds to an $r \times r$ GUE type
matrix with joint distribution of elements proportional to $\exp(- \sigma^{-2} {\rm Tr} \, X^2)$.
We find that indeed (\ref{GH2}) with $N=0$ reclaims (\ref{B}) with $N \mapsto r$,
$x_j^{(N)} \mapsto x_j^{(r)}/\sigma$ as required. Furthermore, with $r=0$ we should reclaim what
we started with --- the $N \times N$ GUE. Using the general formula
$$
\lim_{r \to 0^+} {1 \over (r-1)!} \int_0^x (x - u)^{r-1} f(u) \, du = f(x)
$$
we can check that (\ref{GH2}) so simplifies.

The case $\sigma = 1$ corresponds to an additive finite rank perturbation of an $(N+r) \times (N+r)$ GUE matrix.
Thus extending the remark below (\ref{1})
we then have $G_{N+r} = X + \mu {\rm diag} \, [1^r,0^N]$ 
The eigenvalue PDF in this case can be evaluated using the well known
Harish-Chandra/Itzykson-Zuber integral \cite{Pe06,DF06}. First, to make sense of (\ref{GH2}) in the case
$\sigma = 1$, we must subtract a multiple of the final column from the first $N$ columns to replace the
lower terminal of integration therein by 0 (otherwise the integral is not convergent). Then setting
\begin{equation}\label{pk}
p_{k-1}(u) = \Big ( - {1 \over 2 \mu} \Big )^{r} e^{2 \mu u} {d^{r} \over d u^{r} }
e^{-2 \mu u} u^{k-1},
\end{equation}
and integrating by parts shows
$$
{1 \over (r-1)!} \int_0^x (x - u)^{r-1} e^{-2 \mu u} p_{k-1}(u) \, du =
\Big ( {1 \over 2 \mu} \Big )^{r} (e^{-2\mu x} x^{k-1} - \delta_{k,1} ).
$$
It follows that, up to proportionality, (\ref{GH2}) simplifies to
\begin{eqnarray*}
&& e^{-\sum_{j=1}^{N+r} (\lambda_j^{(N+r)})^2)} \det [ q_{k-1}(\lambda_j^{(N+r)}]_{j,k=1,\dots,N+r}
\nonumber \\
&& \quad \times
\det \Big [ [(\lambda_j^{(N+r)})^{k-1} ]_{j=1,\dots,N+r \atop k=1,\dots,r} \quad
[ e^{2 \mu \lambda_j^{(N+r)} } (\lambda_j^{(N+r)})^{r-s} ]_{j=1,\dots,N+r \atop s=1,\dots,r} \Big ],
\end{eqnarray*}
in keeping with the known expression.

\section{Correlations for $r=1$}
\subsection{The correlation kernel}
The simplest case of (\ref{GH2}) beyond the GUE itself is $r=1$. Here we take up the task of calculating
the corresponding correlation functions. Our strategy is to seek a choice of $\{q_j(\lambda)\}$,
$\{p_j(u)\}$ such that the structure of a biorthogonal ensemble \cite{Mu96,Bor99}, \cite[Section 5.8]{Fo10}
\begin{equation}\label{4.1}
{1 \over \bar{C}_N} \prod_{l=1}^N w_2(\lambda_l) \det [\xi_j(\lambda_k) ]_{j,k=1,\dots,N+1}
\det [ \eta_j(\lambda_k) ]_{j,k=1,\dots,N+1}.
\end{equation}
 Once this is achieved, the general $k$-point correlation is given in terms
of a $k \times k$ determinant according to
\begin{equation}\label{4.3}
\rho_{(k)}(\lambda_1,\dots,\lambda_k) = 
\det [ K_{N+1}(\lambda_j, \lambda_l) ]_{j,l=1,\dots,k}
\end{equation}
where the so called correlation kernel $K_{N+1}$ is given by
\begin{equation}\label{CK}
K_{N+1}(\lambda,\nu) := (w_2(\lambda)w_2(\nu))^{1/2}\sum_{p,q=1}^{N+1} 
({\bf g}_{N+1}^{-1})_{p,q}\xi_p(\lambda) \eta_q(\nu) .
\end{equation}
In (\ref{CK}) ${\bf g}_{N+1}^{-1}$ refers to the inverse of the $(N+1) \times (N+1)$ matrix
of inner products
\begin{equation}\label{4.2}
{\bf g}_{N+1} = \Big [ \int_{-\infty}^\infty w_2(\lambda) \xi_j(\lambda) \eta_k(\lambda) \, d \lambda
\Big ]_{j,k=1,\dots,N+1}. 
\end{equation}
We will seek a form (\ref{4.1}) so that the inverse of (\ref{4.2}) is simple to compute.

\begin{prop}
Let $\{H_p(x)\}$ denote the Hermite polynomials, characterized by the orthogonality
\begin{equation}\label{4.6a}
\int_{-\infty}^\infty e^{-x^2} H_j(x) H_k(x) \, dx = {\mathcal N}_j \delta_{j,k}, \qquad
{\mathcal N}_j = 2^j j! \sqrt{\pi}.
\end{equation}
Introduce $\{\alpha_p\}$, $\{\beta_p\}$ as expansion coefficients
\begin{eqnarray}\label{4.6b}
e^{(1-\sigma^{-2}) x^2 + 2 \mu \sigma^{-2} x} \int_0^x
e^{(-1 + \sigma^{-2}) u^2 - 2 \mu \sigma^{-2} u} \, du & = & \sum_{p=0}^\infty \alpha_p H_p(x) \nonumber \\
 e^{(1 - \sigma^{-2})x^2 +2 \mu \sigma^{-2} x} & = & \sum_{p=0}^\infty \beta_p H_p(x).
\end{eqnarray}
Write
\begin{equation}\label{4.6c}
e^{-(x^2 + y^2)/2} \sum_{j=1}^{N-1} {1 \over {\mathcal N}_{j-1} } H_{j-1}(x) H_{j-1}(y) =:
K_{N-1}^{\rm GUE}(x,y)
\end{equation}
(this is the correlation kernel for the $(N-1) \times (N-1)$ GUE)
and set
\begin{equation}\label{4.6d}
\eta_j(x) = \left \{ \begin{array}{ll} \sum_{p=N-1}^\infty \beta_p H_p(x), & j=N \\[.2cm]
 \sum_{p=N-1}^\infty \alpha_p  H_p(x), & j=N+1 \end{array} \right.
\end{equation}
The correlation kernel (\ref{CK}) for the $r=1$ case of (\ref{GH2}) is given by
\begin{eqnarray}\label{4.6e}
K_{N+1}(x,y)  =  K_{N-1}^{\rm GUE}(x,y) + {e^{-(x^2 + y^2)/2} \over \beta_{N-1} \alpha_N - \beta_N 
\alpha_{N-1} } \bigg ( \Big ( {\alpha_N \over  \mathcal{N}_{N-1} } H_{N-1}(x)
- {\alpha_{N-1} \over \mathcal{N}_N } H_{N}(x) \Big ) \eta_{N}(y)
 \nonumber \\
 + \Big ( {\beta_{N-1} \over {\mathcal N}_N } H_N(x) -
{\beta_N \over {\mathcal N}_{N - 1} } H_{N-1}(x) \Big )  \eta_{N + 1} (y) \bigg ) 
\end{eqnarray}
\end{prop}

\noindent
{\it Proof.} \quad
For notational convenience write
$$
z(u) = e^{(-1 + \sigma^{-2}) u^2 - 2 \mu \sigma^{-2} u}.
$$
In terms of this notation, specify the monic polynomials $\{p_j(x)\}$  in (\ref{GH2}) according to
\begin{equation}\label{pp}
p_j(x) = - {2^{-j} \over (1 - \sigma^{-2})} {1 \over z(x)} {d \over d x}
\Big ( z(x) H_{j-1}(x) \Big ) \quad (j=1,\dots,N), \qquad p_0(x) = 1
\end{equation}
(cf.~(\ref{pk})). Furthermore, specify the monic polynomials $\{q_j(x)\}$  by
$$
q_j(x) = 2^{-j} H_j(x) \qquad (j=0,\dots,N).
$$

Noting that the first of the specifications (\ref{pp}) implies
$$
\int_{-\infty}^{\lambda_j} z(u) p_k(u) \, du = -
{2^{-j} \over (1 - \sigma^{-2}) } z(\lambda_j) H_{k-1}(\lambda_j) \quad (k=1,\dots,N)
$$
we see that (\ref{GH2}) with the above simplifications and $r=1$ is 
proportional to
\begin{eqnarray}\label{pp1}
&& {1 \over \bar{C}_N }
e^{- \sum_{j=1}^{N+1} \lambda_j^2 } \det [ H_{k-1}(\lambda_j) ]_{j,k=1,\dots,N+r} \nonumber \\
&& \quad \times
\det \bigg [ \Big [ {1 \over z(\lambda_j)} \int_{-\infty}^{\lambda_j} z(u) \, du \Big ]_{j=1,\dots,N+1} \quad
[H_{k-1}(\lambda_j)]_{j=1,\dots,N+1 \atop k=1,\dots,N-1} \quad
\Big [{1 \over z(\lambda_j) } \Big ]_{j=1,\dots,N+1} \bigg ]
\end{eqnarray}
where,  with $C_N$ is specified by (\ref{CN}),
$$
\tilde{C}_N = C_N 2^{N^2 - N} \sigma^{1 + 2N} (\sigma^{-2} - 1)^{N-1} \pi^{-1/2}
$$
and
we have set $\lambda_j^{(N+1)} = \lambda_j$.

The expression (\ref{pp1}) is the form (\ref{4.1}) of a biorthogonal ensemble with
\begin{eqnarray*}
\xi_j(x) & = & H_{j-1}(x) \\[.15cm]
\eta_j(x)  & = & \left \{ \begin{array}{cc} H_{j-1}(x), & j=1,\dots,N-1 \\[.15cm]
\sum_{p=N-1}^\infty \beta_p H_p(x), & j = N \\[.15cm]
\sum_{p=N-1}^\infty \alpha_p  H_p(x), & j = N+1
\end{array} \right. 
\end{eqnarray*}
and ${\bf g}_{N+1}$ a diagonal matrix for rows $j=1,\dots,N-1$ with entries ${\mathcal N}_j$, and a
$2 \times 2$ block diagonal matrix with respect to the last two rows and columns, with $2 \times 2$ block
\begin{equation}\label{3.11a}
\begin{bmatrix} \beta_{N-1} & \alpha_{N-1} \\
\beta_N & \alpha_N \end{bmatrix}.
\end{equation}
It is thus straightforward to calculate ${\bf g}_{N+1}^{-1}$, which when substituted in (\ref{CK}) gives
(\ref{4.6e}). \hfill $\square$

\subsection{The expansion coefficients}
We seek the evaluation of the expansion coefficients in (\ref{4.6b}). Since
\begin{eqnarray}
{\mathcal N}_p \alpha_p & = & \int_{-\infty}^\infty dx \, e^{-\sigma^{-2} x^2 + 2 \mu \sigma^{-2} x}
H_p(x) \int_0^x du \, e^{(-1+\sigma^{-2}) u^2 - 2 \mu \sigma^{-2} u},  \label{na} \\
{\mathcal N}_p \beta_p & = & \int_{-\infty}^\infty  e^{-\sigma^{-2} x^2 + 2 \mu \sigma^{-2} x}
H_p(x) \, dx \label{nb}
\end{eqnarray}
this requires us computing some integrals. 

\begin{prop}
We have
\begin{eqnarray}\label{4b}
&&\mathcal N_p \beta_p = e^{(\mu/\sigma)^2} (1 - \sigma^2)^{p/2} H_p(\mu(1 - \sigma^2)^{-1/2}), \\
&&\mathcal N_p \alpha_p = 
c_1 (1 - \sigma^2)^{p/2}  H_p(\mu(1 - \sigma^2)^{-1/2}) + c_2 
 (1 - \sigma^2)^{p/2} h_p(\mu  (1 - \sigma^2)^{-1/2}), \label{4a}
\end{eqnarray}
where
$$
h_p(x) := \int_{-\infty}^\infty {e^{-u^2} H_p(u) \over x - u} \, du
$$
and
\begin{eqnarray*}
&&c_1 = {\mathcal N}_0 \alpha_0 - {h_0(\mu(1 - \sigma^2)^{-1/2}) \sigma^2 \over
(1 - \sigma^2)^{1/2}) h_1(\mu (1 - \sigma^2)^{1/2})  - 2 \mu h_0(\mu(1 - \sigma^2)^{-1/2})}, \\
&& 
c_2 = {\sigma^2 \over (1 - \sigma^2)^{1/2}) h_1(\mu (1 - \sigma^2)^{-1/2})  - 2 \mu h_0(\mu(1 - \sigma^2)^{-1/2})}.
\end{eqnarray*}
In the case $\mu = 0$ these simplify to give
\begin{eqnarray}
&& {\mathcal N}_{2p} \alpha_{2p} = 0, \qquad  {\mathcal N}_{2p+1} \beta_{2p+1} = 0, \label{1a} \\
&&  {\mathcal N}_{2p+1} \alpha_{2p+1} = \sqrt{\pi} 2^{2p} p! \sigma^2 (\sigma^2 - 1)^p, \label{1b} \\
 && {\mathcal N}_{2p} \beta_{2p} = \sqrt{\pi} {(2p)! \over p!}  \sigma (\sigma^2 - 1)^p.  \label{1c}
\end{eqnarray}
\end{prop}

\noindent
{\it Proof.} \quad 
The evaluation (\ref{4b}) follows immediately from the well known, and readily verified, integral identity
$$
\int_{-\infty}^\infty e^{-(x - y)^2/2u} H_n(x) \, dx =
(2 \pi u)^{1/2} (1 - 2u)^{n/2} H_n(y(1 - 2u)^{-1/2}).
$$
Alternatively, we can make use of the three term recurrence
\begin{equation}\label{Hp}
H_{p+1}(x) = 2x H_{p}(x) - 2p H_{p-1}(x) \qquad (p=0,1,\dots)
\end{equation}
and the differentiation formula
\begin{equation}\label{dt}
{d \over dx} H_{p}(x) = 2p H_{p-1}(x)
\end{equation}
to deduce the recurrence
\begin{equation}\label{Hp1a}
{\mathcal N}_p \beta_p = 2 \mu {\mathcal N}_{p-1} \beta_{p-1} + 2 (p-1) (\sigma^2 - 1)
 {\mathcal N}_{p-2} \beta_{p-2} \qquad (p=1,2,\dots).
\end{equation}
With ${\mathcal N}_p \beta_p = (1 - \sigma^2)^{p/2} y_p$ this reads
\begin{equation}\label{3.22}
y_p = 2 \mu (1 - \sigma^2)^{-1/2} y_{p-1} - 2(p-1) y_{p-2}.
\end{equation}
Now (\ref{3.22}) with $p \mapsto p+1$ and $x = \mu (1 - \sigma^2)^{-1/2}$ is identical to (\ref{Hp}).
This establishes (\ref{4b}) up to the value of $\mathcal N_0 \beta_0$, which can be checked directly.

The advantage in setting up a recurrence is that the same strategy works for the integral
(\ref{nb}). Thus making use of (\ref{Hp}) and (\ref{dt}) and integration by parts we deduce
\begin{equation}\label{pap}
{\mathcal N}_p \alpha_p = 2 \mu {\mathcal N}_{p-1} \alpha_{p-1} + 2 (p-1) (\sigma^2 - 1)
 {\mathcal N}_{p-2} \alpha_{p-2} + \sigma^2 \sqrt{\pi} \delta_{p,1} \qquad (p=1,2,\dots),
\end{equation}
This with ${\mathcal N}_p \alpha_p = (1 - \sigma^2)^{p/2} x_p$ reads
$$
x_p = 2 \mu (1 - \sigma^2)^{-1/2} x_{p-1} - 2 (p-1) x_{p-2} + \sigma^2 (1 - \sigma^2)^{-1/2}
\delta_{p,1}.
$$
Hence $\{x_p\}_{p=2,3,\dots}$ satisfies the recurrence (\ref{Hp}), with $x = \mu (1 - \sigma^2)^{-1/2}$ and
for $p=1,2,\dots$. The recurrence, being of second order, has two linearly independent solutions
$H_p(x)$ and $h_p(x)$ so we have
\begin{equation}\label{pap1}
{\mathcal N}_p \alpha_p = c_1 (1 - \sigma^2)^{p/2} H_p(\mu(1 - \sigma^2)^{-1/2}) +
c_2 (1 - \sigma^2)^{p/2}  h_p(\mu(1 - \sigma^2)^{-1/2}).
\end{equation}
The values of $c_1$ and $c_2$ follow by comparing (\ref{pap}) in the case $p=1$.

In the case $\mu = 0$ (\ref{Hp1a}) and (\ref{pap})
reduce to first order recurrences. Their solutions imply the more explicit formulas (\ref{1a})--(\ref{1b}). 
\hfill $\square$

\subsection{The case $\sigma = 1$}
As revised below (\ref{1}), the case $\sigma = 1$ corresponds to the additive rank 1 perturbation
$X \mapsto X + (\mu/(N+1)) \vec{e}_1 \vec{e}_1^T$ where $X$ is a member of the $(N+1) \times (N+1)$
GUE. The correlation kernel for such random matrices has previously been expressed in the form \cite{DF06}
\begin{equation}\label{Ge}
K_{N+1}(x,y) = K_N^{\rm GUE}(x,y) + {(-1)^N \over \sqrt{\pi} } e^{-x^2/2 - y^2/2} H_N(x)
\tilde{\Gamma}_{N+1}(\mu;y)
\end{equation}
where
\begin{equation}\label{Gg}
\tilde{\Gamma}_{N+1}(\mu;y) := \int_{\mathcal C_{\{0,-2\mu\}}}
{e^{-yz - z^2/4}\over z^N (z+ 2 \mu) } \,
{dz \over 2 \pi i}.
\end{equation}
Here $\mathcal C_{\{0,-2\mu\}}$ is a simple closed contour encircling 0 and $-2\mu$.
Furthermore, we know (\ref{Gg}) has the expansion \cite{BFF09}
\begin{equation}\label{Gg1}
\tilde{\Gamma}_{N+1}(\mu;y) = (-1)^N \bigg ( {e^{2 \mu y - \mu^2} \over (2 \mu)^N } -
\sum_{j=0}^{N-1} {1 \over (2 \mu)^{p+1} } {H_{N-1-p}(y) \over 2^{N-1-p} (N-1-p)! } \bigg ).
\end{equation}

To deduce (\ref{Ge}) from (\ref{4.6e}) we first rearrange the latter to read
\begin{eqnarray}\label{9}
K_{N+1}(x,y) & = & K_{N-1}^{\rm GUE}(x,y)  \nonumber \\
&&  + e^{- (x^2 + y^2)/2} \Big ( H_{N-1}(x) \sum_{p=N-1}^\infty {\alpha_N \beta_p - \beta_N \alpha_p \over
\mathcal N_{N-1} (\alpha_N \beta_{N-1} - \alpha_{N-1} \beta_N) } H_p(y)   \nonumber \\
 && + H_{N}(x) \sum_{p=N}^\infty {\beta_{N-1} \beta_p - \alpha_{N-1} \beta_p \over
\mathcal N_{N} (\alpha_N \beta_{N-1} - \alpha_{N-1} \beta_N) } H_p(y) \Big ).
\end{eqnarray}

\begin{lemma}
In the limit $\sigma \to 1$
\begin{eqnarray*}
{\mathcal N_p \over \mathcal N_{N-1}} {\alpha_N \beta_p - \beta_N \alpha_p \over
\alpha_N \beta_{N-1} - \beta_N \alpha_{N-1} } & \to &
\left \{ \begin{array}{ll}1, & p=N-1 \\
0, & p \ge N, \end{array} \right. \\
{\mathcal N_p \over \mathcal N_N} {\alpha_{N-1} \beta_p - \beta_{N-1} \alpha_p \over
\alpha_N \beta_{N-1} - \beta_N \alpha_{N-1} } & \to & (2 \mu)^{p-N}, \qquad p \ge N.
\end{eqnarray*}
\end{lemma}

\noindent
{\it Proof.} \quad These results follow by applying the recurrences (\ref{Hp1a}) and
(\ref{pap}) to the numerators. \hfill $\square$

\medskip
\begin{cor}\label{c5}
In the limit $\sigma \to 1$
\begin{eqnarray}\label{Sp}
K_{N+1}(x,y) & = & K_{N}^{\rm GUE}(x,y) + e^{-(x^2 + y^2)/2} H_N(x) \sum_{p=N}^\infty
(2 \mu)^{p-N} {H_p(y) \over \mathcal N_p} \nonumber \\
& = &  K_{N}^{\rm GUE}(x,y) + e^{- (x^2 + y^2)/2} {H_N(x) \over (2 \mu)^N }
\Big ( e^{2 \mu y - \mu^2} - \sum_{p=0}^{N-1} (2 \mu)^p {H_p(y) \over \mathcal N_p} \Big ),
\end{eqnarray}
where the second equality follows by taking $\sigma \to 1$ in (\ref{4b}) and substituting in 
the second equation of (\ref{4.6b}).
\end{cor}

Recalling (\ref{Gg1}) we see that (\ref{Sp}) is identical to the known result (\ref{Ge}).

\subsection{The case $\mu = 0$}
Substituting (\ref{1a})--(\ref{1c}) in (\ref{4.6d}) and (\ref{4.6e}) we obtain the
specialization of the correlation kernel in the case $\mu = 0$.

\begin{prop}\label{p6}
Suppose $\mu = 0$. For $N$ even
\begin{eqnarray}
K_{N+1}(x,y) & =  &  K_{N-1}^{\rm GUE}(x,y) \nonumber \\
&& + e^{- (x^2 + y^2)/2 } \bigg (
{H_{N-1}(x) \over 2^{N-2} (N/2 - 1)! (\sigma^2 - 1)^{N/2 - 1} }
\sum_{p=N/2 - 1}^\infty { 2^{2p} p! (\sigma^2 - 1)^p  H_{2p+1}(y)\over \mathcal N_{2p+1} }   
\nonumber \\
&& \vspace{3cm} 
+ {H_N(x) (N/2)! \over N! (\sigma^2 - 1)^{N/2} }
\sum_{p=N/2}^\infty {1 \over \mathcal N_{2p} } {(2p)! \over p!} (\sigma^2 - 1)^p H_{2p}(y) \bigg ),
\end{eqnarray}
while for $N$ odd
\begin{eqnarray}
K_{N+1}(x,y) & =  &  K_{N-1}^{\rm GUE}(x,y) \nonumber \\
&& + e^{- (x^2 + y^2)/2 } \bigg (
{H_{N-1}(x) ((N-1)/2)! \over  (N - 1)! (\sigma^2 - 1)^{(N - 1)/2} }
\sum_{p=(N - 1)/2}^\infty { (2p)! (\sigma^2 - 1)^p  H_{2p}(y)\over p! \mathcal N_{2p} }   
\nonumber \\
&& \quad
+ {H_N(x)  \over 2^{N-1} ((N-1)/2)! (\sigma^2 - 1)^{(N-1)/2} }
\sum_{p=(N-1)/2}^\infty {1 \over \mathcal N_{2p+1} } 2^{2p} p! (\sigma^2 - 1)^p H_{2p+1}(y) \bigg ).
\end{eqnarray}

\end{prop}

\section{Soft edge scaling and the phase transition}
\subsection{Secular equation}
As revised in the Introduction, rank 1 perturbations can lead to a phase transition
with respect to the location of the largest eigenvalue. In \cite{BFF09} we showed how
for the additive rank 1 perturbation the corresponding secular equation associated with
the eigenvalue problem  could be used to predict the critical value
$\varepsilon = 1/\sqrt{2N}$ (or equivalently $\mu = \sqrt{N/2}$) at which separation occurs, as well as the
formula (\ref{c}) for the location of the largest eigenvalue in the separated phase.
Here we will repeat those considerations in the case of the modified
$(N+1) \times (N+1)$ GUE matrices as specified by (\ref{1}) (the same reasoning applies
to the GOE version of this construction).

The secular equation determining the eigenvalues in this case is the condition (\ref{3.1}) with
$\{a_j\}$ the eigenvalues of an $N \times N$ GUE matrix.
Averaging over $|v_N|^2$ and $c_N$ it reduces to
\begin{equation}\label{la}
\lambda - \mu = {\sigma^2 \over 2} \sum_{j=1}^N {1 \over \lambda - a_j}.
\end{equation}
We seek the range of values of $\sigma^2$ which
permit this equation to be solved for $|\lambda | > \sqrt{2N}$, and the corresponding
value of $|\lambda|$. In view of the interlacing (\ref{4}) this will correspond to the
eigenvalues at the bottom and top of the spectrum separating from the
GUE spectrum. We know the latter, to leading order, has density given by the Wigner
semi-circle law
\begin{equation}\label{5.1}
\rho_{(1)}^{\rm W}(\lambda) = {\sqrt{2N} \over \pi} \sqrt{1 - {\lambda^2 \over 2N} },
\qquad  |\lambda| < 2N.
\end{equation}
But in general, for large $N$ with $\{a_j\}$ forming a continuum supported on
$I$ with density $\rho_{(1)}(y)$,
\begin{equation}\label{5.1a}
 \sum_{j=1}^N {1 \over \lambda - a_j} \sim \int_I {\rho_{(1)}(y) \over \lambda - y} \, dy.
\end{equation}
Substituting (\ref{5.1}) in (\ref{5.1a}) the resulting integral can be evaluated
(see e.g.~ \cite[eq.~(2.10)]{BFF09}). Thus
$$
\int_{- \sqrt{2N}}^{\sqrt{2N}} {\rho_{(1)}^{\rm W}(y) \over \lambda - y} \, dy =
\lambda \Big ( 1 -  \sqrt{ 1 - {\lambda^2 \over 2N} } \Big ), \qquad  |\lambda| > 2N
$$
and so in the large $N$ limit the averaged secular equation (\ref{1a}) gives that for
eigenvalue separation we must have
\begin{equation}\label{co}
\lambda - \mu = \lambda {\sigma^2 \over 2}  \Big ( 1 -  \sqrt{ 1 - {\lambda^2 \over 2N} } \Big ) 
, \qquad  |\lambda| > 2N.
\end{equation}

Writing 
\begin{equation}\label{4.4a}
\mu = c \sqrt{N/2}
\end{equation}
 and substituting $\lambda = \sqrt{2N}$ --- the largest value of $\lambda$ which does
not correspond to separation --- we deduce that separation of the largest eigenvalue occurs for
\begin{equation}\label{co1}
\sigma^2 + c > 2.
\end{equation}
Similarly, we deduce that separation of the smallest eigenvalue occurs for
\begin{equation}\label{co2}
\sigma^2 - c > 2.
\end{equation}
And solving (\ref{co}) under the conditions (\ref{co1}) and (\ref{co2}) gives that the corresponding location
of the separated eigenvalues will be at
\begin{equation}\label{co3}
\lambda = \sqrt{N \over 2}
{\sigma^4 + c^2 \over (1 - \sigma^2/2) c \pm \sigma^2 ( (c/2)^2 - (1 - \sigma^2) )^{1/2} }
\end{equation}
with the choice of $\pm$ corresponding to (\ref{co1}) and (\ref{co2}) respectively. Note that when
$\sigma = 1$ this reclaims (\ref{c}), while for $c=0$ it reads
\begin{equation}\label{co3a}
\lambda = \pm \sqrt{N \over 2}
{\sigma^2  \over  ( \sigma^2 -1 )^{1/2} }
\end{equation}

\subsection{The case $\sigma = 1$}
The correlation kernel in this case is given by (\ref{Sp}) or equivalently (\ref{Ge}).
We know from (\ref{c}) that $\mu = \sqrt{N/2}$ is the critical value of $\mu$ in relation to
separation of the largest eigenvalue from the Wigner semi-circle. 
A critical region occurs when the eigenvalues are scaled about the neighbourhood
of the edge of the Wigner circle so that their spacing is O$(1)$, and the mean $\mu$ is
scaled about the critical value $\sqrt{N/2}$. Explicitly, in (\ref{Sp}) this is achieved by
introducing the scaling variables $X,Y,s$ according to
\begin{equation}\label{xy}
x = \sqrt{2N} + {X \over 2^{1/2} N^{1/6}}, \qquad
y = \sqrt{2N} + {Y \over 2^{1/2} N^{1/6}}, \qquad
\mu = \sqrt{N/2}\Big ( 1 + {s \over N^{1/3}} \Big ).
\end{equation}
The task then is to compute the scaled $N \to \infty$ limit, which we know from
\cite{Pe06,DF06} gives
\begin{equation}\label{KA}
\lim_{N \to \infty} {1 \over \sqrt{2} N^{1/6}} K_{N+1}(x,y) =
K^{\rm soft}(X,Y) + {\rm Ai}(X) \int_{-\infty}^Y e^{-s(Y-t)} {\rm Ai}(t) \, dt
\end{equation}
where $K^{\rm soft}(X,Y)$ denotes the Airy kernel \cite{Fo93a}. 
$$
K^{\rm soft}(X,Y) = {{\rm Ai}(X) {\rm Ai}'(Y) - {\rm Ai}(Y) {\rm Ai}'(X) \over X - Y }.
$$
 As written the integral in
(\ref{KA}) requires $s \ge 0$ to be convergent, 
Noting that
\begin{eqnarray}
 \int_{-\infty}^Y e^{-s(Y-t)} {\rm Ai}(t) \, dt  & = &  \int_{-\infty}^\infty e^{-s(Y-t)} {\rm Ai}(t) \, dt -
 \int_{Y}^\infty e^{-s(Y-t)} {\rm Ai}(t) \, dt \nonumber \\
 & = & e^{-sY + Y^3/3} -  \int_{Y}^\infty e^{-s(Y-t)} {\rm Ai}(t) \, dt
\end{eqnarray}
gives this meaning for general $s$.
The scaled correlation kernel
(\ref{KA}) is precisely that which is found for the analogous scaling limit of the correlation
kernel of the
$r=1$ spiked complex Wishart ensemble \cite{BBP05}.

\subsection{The case $\mu = 0$} 
The sums in Proposition \ref{p6} are very similar to those appearing in an earlier work \cite{Fo99b}.
There the main tool in the analysis of the soft edge limit of the sums is the uniform asymptotic expansion
\begin{equation}\label{HA}
{e^{-x^2/2} H_n(x) \over \mathcal N_n^{1/2} } \Big |_{x = (2n)^{1/2} - u/2^{1/2} n^{1/6} } =
2^{1/4} n^{-1/12} \Big ( {\rm Ai}(-u) + {\rm O}(n^{-2/3}) \Big )
\end{equation}

\begin{prop}\label{p7}
Set 
$$
\sigma^2 = 2 - 2s/N^{1/3}
$$
and scale $x$ and $y$ as in (\ref{xy}). Then (\ref{KA}) again results from the scaling limit of both the
$N$ even and $N$ odd correlation kernels in Proposition \ref{p6}.
\end{prop}

\noindent
{\bf Proof.} \quad Consider for definiteness the case $N$ even. The terms in addition to $K_{N-1}^{\rm GUE}(x,y)$
of the correlation kernel can be written
$$
e^{-(x^2+y^2)/2} \bigg (
{H_{N-1}(x) \over ({\mathcal N}_{N-1})^{1/2} } \sum_{p=0}^\infty a_{p,N} (\sigma^2 - 1)^p
{H_{N-1+2p}(y) \over  {\mathcal N}_{N-1+2p}^{1/2} } +
 {H_{N}(x) \over {\mathcal N}_{N}^{1/2} }\sum_{p=0}^\infty b_{p,N} (\sigma^2 - 1)^p
{H_{N+2p}(y) \over  {\mathcal N}_{N+2p}^{1/2} } \bigg )
$$
where
$$
a_{p,N} = \Big ( {\mathcal N_{N-1} \over \mathcal N_{N-1+2p} } \Big )^{1/2}
{2^{2p} (N/2 - 1 + p)! \over (N/2 - 1)! }, \qquad
b_{p,N} = \Big ( {\mathcal N_{N} \over \mathcal N_{N+2p} } \Big )^{1/2}
{ (N/2)! (N + 2p)! \over (N/2 +p)! N! }.
$$
For large $N$ and $p \ll N$ we see from Stirling's formula that $a_{p,N}$ and $b_{p,N}$
tend to 1. The key remaining step is to use (\ref{HA}) in the summands. Considering the first
of these, for definiteness, we substitute (\ref{HA}) with $n= N + 2p-1$ and $-u = Y - 2p/N^{1/3}$.
Writing $2p/N^{1/3} = t$ we see that a Riemann integral approximation to the sum results, giving
that the large $N$ form of the latter is
\begin{equation}\label{k1}
{N^{1/4} \over 2^{3/4}} \int_0^\infty e^{-st}  {\rm Ai}(Y - t) \, dt.
\end{equation}
Also, a literal application of (\ref{HA}) shows that the leading asymptotic form of the term outside
the first sum is equal to
\begin{equation}\label{k2}
2^{1/4} N^{-1/12} {\rm Ai} (X).
\end{equation}
The second term gives the same leading asymptotics. Thus we have to multiply twice (\ref{k1}) with (\ref{k2}),
and divide by $2^{1/2} N^{1/6}$ (recall the LHS of (\ref{KA}) in relation to this factor). The correction term to
$K^{\rm soft}(X,Y)$ on the RHS of  (\ref{KA}) results. \hfill $\square$

\subsection{General $\sigma^2$, $\mu > 0$ about eigenvalue separation}
The correlation kernel (\ref{9}) consists of the GUE kernel and a correction term, the latter
involving a summation over Hermite polynomials. The cases $\sigma^2 = 1$ and $\mu = 0$ have tbe
special feature that the coefficients exhibit simple functional forms (Corollary
\ref{c5} and Proposition \ref{p6}). In contrast, for general $\sigma$ and $\mu$ the coefficients are
given in terms of Hermite polynomials evaluated at a special point, and a certain Hilbert transformation
of the same Hermite polynomial evaluated at this point. The key to calculating the large $N$, $p$, $p \ll N$
form of the coefficients in this case relies on their characterization in terms of the solution of
a difference equation, or more explicitly that of the scaled coefficients
\begin{equation}\label{ga1}
\Big ( {{\mathcal N}_{N-1} \over {\mathcal N}_{N-1+p} } \Big )^{1/2}
{ \tilde{\alpha}_N \tilde{\beta}_{N-1+p} - \tilde{\beta}_N \tilde{\alpha}_{N-1+p} \over
 \tilde{\alpha}_N \tilde{\beta}_{N-1} - \tilde{\alpha}_{N-1} \tilde{\beta}_{N} } =: \gamma_p^{(1)}
\end{equation}
and
\begin{equation}\label{ga2}
\Big ( {{\mathcal N}_{N} \over {\mathcal N}_{N-1+p} } \Big )^{1/2}
{ \tilde{\beta}_N \tilde{\alpha}_{N-1+p} - \tilde{\alpha}_N \tilde{\beta}_{N-1+p} \over
 \tilde{\beta}_N \tilde{\alpha}_{N-1} - \tilde{\beta}_{N-1} \tilde{\alpha}_{N} } =: \gamma_p^{(2)},
\end{equation}
where $\tilde{\alpha}_p = \mathcal N_p \alpha_p$, $\tilde{\beta}_p = \mathcal N_p \beta_p$.

\begin{prop}
With $\mu$ replaced in favour of $c$ according to (\ref{4.4a}), and with
\begin{equation}\label{2x}
x_\pm := {c \pm (c^2 - 4 (1 - \sigma^2) )^{1/2} \over 2}
\end{equation}
for $N \to \infty$ we have
\begin{equation}\label{2y}
\gamma_p^{(1)} = {x_+^p x_- - x_+ x_-^p \over x_- - x_+}, \qquad
\gamma_p^{(2)} = {x_+^p  - x_-^p \over x_+ - x_-}.
\end{equation}
\end{prop}

\noindent
{\it Proof.} \quad We know that both $\tilde{\alpha}_p$ and  $\tilde{\beta}_p$ satisfy the
same recurrence (\ref{Hp1a}). But up to the scale factor $(\mathcal N_{N-1+p})^{1/2}$,
$\gamma_p^{(1)}$ and  $\gamma_p^{(2)}$ are linear combinations of $\tilde{\alpha}_{N-1+p}$ and
$\tilde{\beta}_{N-1+p}$. Thus in (\ref{Hp1a}), by replacing $p$ by $N-1+p$, multiplying by through by
$(\mathcal N_{N-1+p})^{1/2}$ and making the substitution (\ref{4.4a})
we conclude that in the limit $N \to \infty$ both
$\gamma_p^{(1)}$ and  $\gamma_p^{(2)}$ satisfy the constant coefficient recurrence
$$
\gamma_{q+1} = c \gamma_q + (\sigma^2 - 1) \gamma_{q-1} \qquad (q=1,2,\dots).
$$
For $\gamma_p^{(1)}$ this is to be solved subject to the initial condition
$\gamma_0$, $\gamma_1 = 0$, while for $\gamma_p^{(2)}$ it is subject to the initial
condition $\gamma_0 = 0$, $\gamma_1 = 1$. Solving the recurrence gives the stated result.
\hfill $\square$

Let us now write
\begin{equation}\label{3aa}
c = \hat{c} + {s_1 \over N^{1/3} }, \qquad
\sigma^2 = (\hat{\sigma})^2 - {s_2 \over N^{1/3} }
\end{equation}
with
\begin{equation}\label{3b}
(\hat{\sigma})^2 +  \hat{c} = 2.
\end{equation}
According to (\ref{co1}) we are thus perturbing about the critical values for separation
of the largest eigenvalue. Substituting in (\ref{2x}) shows
\begin{equation}\label{4.0}
x_+ \sim (\hat{c} - 1) + {\rm O}\Big ( {1 \over N^{1/3}} \Big ), \qquad
x_- \sim 1 + {(\hat{c}- 3) s_1 + 2 s_2 \over 2 (\hat{c} - 2) N^{1/3} }.
\end{equation}
Restricting attention to the case
$\mu > 0$ so that $\hat{c} > 0$ (recall the case $\hat{c} = 0$ as a special case covered above),
and furthermore, from (\ref{3b}), $ \hat{c} < 2$ (we exclude the case $\hat{c} = 2$ since this
would mean $\hat{\sigma}^2 = 0$ which corresponds to a decoupled eigenvalue),
we observe that $|x_+| < 1$. Recalling (\ref{2y}) this tells us that for $p$ large
$$
\gamma_p^{(1)} \sim - {x_-^p \over \hat{c} - 2}, \qquad
\gamma_p^{(2)} \sim  {(\hat{c} - 1) x_-^p \over \hat{c} - 2 }.
$$
Substituting in (\ref{9}) we therefore have that for $N$, $p$ large, $p \ll N$,
\begin{equation}\label{4.22a}
K_{N+1}(x,y) \sim K_{N-1}^{\rm GUE}(x,y) + e^{- (x^2 + y^2)/2} {H_N(x) \over \mathcal N_N^{1/2} }
\sum_{p=0}^\infty x_-^p {H_{p+N}(y) \over \mathcal N_{p+N}^{1/2} }.
\end{equation}
But this is precisely the setting of the sums in the proof of Proposition \ref{p7}, which we know
give rise to the second term in (\ref{KA}).

\begin{prop}
Let $c$ be related to $\mu$ by (\ref{4.4a}), scale $c$ and $\sigma^2$ according to (\ref{3aa}) with
condition (\ref{3b}). Furthermore specify $s$ by
\begin{equation}\label{4.22b}
s = {(\hat{c} - 3) s_1 + 2 s_2 \over 2 (2 - \hat{c}) }.
\end{equation}
One has that the limiting correlation kernel is again given by (\ref{KA}).
\end{prop}

For $\hat{c} < 0$ we see from (\ref{4.0}) that $x_+ < -1$. One is then faced with a formally
divergent sum
\begin{equation}\label{5}
\sum_{p=0}^\infty (-1)^p |x_+|^p {H_{p+N}(y) \over {\mathcal N}_{p+N}^{1/2}}, \qquad
|x_+| > 1.
\end{equation}
Our method of analysis thus breaks down, 
whereby it was required that the limiting form of the coefficients be used inside of the summation.
An alternative approach, with the sum  rewritten --- perhaps as an integral --- is called for, although
this is yet to be found.
Nonetheless, with the case $\hat{c} = 0$ also giving (\ref{KA}),
we have no reason to expect anything different for $\hat{c} < 0$, which would be the case if we
could show that the term involving (\ref{5}) vanished in the $N \to \infty$ limit. That the details
of how this comes about are different may be due to the smallest (i.e.~most negative) eigenvalue already
being separated from the bulk of the spectrum in the case $\hat{c} < 0$, but not in the case
$\hat{c} > 0$, an effect which must be encoded in the correction terms to the GUE kernel. 

This detail aside,
our study then indicates that (\ref{KA}) is the universal correlation kernel for the scaling state
about the separation of the largest eigenvalue due to a rank 1 type perturbation, as it is independent of 
the cause of the perturbation being with respect to the mean, the variance, or a combination of both.

In Proposition \ref{p3} of Section 2 the eigenvalue PDF for the $r$-bordered GUE was given.
To proceed as in Section 3 and give the explicit form of the correlation kernel would require inverting
a $2r \times 2r$ matrix (recall (\ref{3.11a})). This in turn puts a stop to us proceeding to
study the scaling limit for general $r$. It is known in the special case $\sigma^2 = 1$ from
\cite{Pe06,DF06}.

It is natural to enquire into the analogous result for bordered GOE matrices. Very recently a method
of analysis involving reduction to tridiagonal form
\cite{DE02} and stochastic differential equations has been introduced to study spiked real
Wishart matrices \cite{BV10}. A bordered GOE matrix can be reduced to
tridiagonal form \cite{FR02b}, and one would expect this method to lead the conclusion that
the eigenvalue separation effect for bordered GOE matrices shares the same universality class as
that for spiked real
Wishart matrices.

\section*{Acknowledgements}
This work was undertaken as part of an ARC International Linkeage Fellowship.
KEB was further supported by the NSF grant \#DMR-0908286  and by the
Texas Advanced Research Program grant \#95921. PJF benefited from discussions with
A.~Bloemendal and B.~Vir\'ag at the AIM workshop `Brownian motion and Random Matrices'
in December 2009.


\providecommand{\bysame}{\leavevmode\hbox to3em{\hrulefill}\thinspace}
\providecommand{\MR}{\relax\ifhmode\unskip\space\fi MR }
\providecommand{\MRhref}[2]{%
  \href{http://www.ams.org/mathscinet-getitem?mr=#1}{#2}
}
\providecommand{\href}[2]{#2}

\end{document}